\documentclass[pra,amsmath,amssymb,amsfonts,nofootinbib]{revtex4}

\usepackage{bm,graphicx,mathrsfs}

\usepackage{amsmath,bbm}
\usepackage{amsfonts,amssymb}
\usepackage{graphicx}
\usepackage{epsfig}
\usepackage{times}
\usepackage{verbatim}
\newcommand{\be}{\begin{equation}}
\newcommand{\ee}{\end{equation}}
\newcommand{\bq}{\begin{eqnarray}}
\newcommand{\eq}{\end{eqnarray}}
\newcommand{\ba}{\begin{align}}
\newcommand{\ea}{\end{align}}

\newcommand{\tr}{{\rm Tr}}
\newcommand{\E}{\mathbb{E}}
\newcommand{\cc}{\mathbb{C}}
\newcommand{\cR}{\mathbb{R}}
\newcommand{\I}{\mathbb {I}}

\newcommand{\raw}{\rightarrow}

\newcommand{\ket}[1]{\left | \, #1 \right\rangle}
\newcommand{\bra}[1]{\left \langle #1 \, \right |}
\newcommand{\braket}[2]{\left\langle\, #1\,|\,#2\,\right\rangle}
\newcommand{\proj}[1]{\ket{#1}\bra{#1}}

\newcommand{\oket}[1]{ | \, #1 \rangle\rangle}
\newcommand{\obra}[1]{ \langle  \langle #1 \,  | }
\newcommand{\obraket}[2]{\langle\langle\, #1\,|\,#2\,\rangle\rangle}
\newcommand{\oproj}[1]{\oket{#1}\obra{#1}}

\newcommand{\bC}{\mathbb{C}}

\newcommand{\cH}{\mathcal{H}}

\newcommand{\cM}{\mathcal M}

\newcommand{\Sp}{\,\,\,\,\,\,}
\newcommand{\no}{\nonumber\\}

\usepackage{color}

\newtheorem{theorem}{Theorem}
\newtheorem{definition}[theorem]{Definition}
\newtheorem{lemma}[theorem]{Lemma}

\newtheorem{corollary}[theorem]{Corollary}

\def\Proof{\noindent\textsc{Proof:}}
\def\proof{\Proof}

\def\qed{\leavevmode\unskip\penalty9999 \hbox{}\nobreak\hfill
     \quad\hbox{\leavevmode  \hbox to.77778em{%
               \hfil\vrule   \vbox to.675em%
               {\hrule width.6em\vfil\hrule}\vrule\hfil}}
     \par\vskip3pt}

\begin{document}

\title{\sc{\Large {Quantum Chi-Squared and Goodness of Fit Testing}}}
\author{Kristan Temme$^{1}$   and  Frank Verstraete$^{2}$ }
\affiliation{$^1$IQIM, California Institute of Technology, Pasadena, California 91125, USA}
\affiliation{$^{2}$ Fakult\"at f\"ur Physik, Universit\"at Wien, Boltzmanngasse 5, 1090 Wien, Austria}
\affiliation{$^{2}$ Faculty of Science, Ghent University, B-9000 Ghent, Belgium}
\date{\today}

\begin{abstract} 
A quantum mechanical hypothesis test is presented for the hypothesis that a certain setup produces a given quantum state. Although the classical and the quantum problem are very much related to each other, the quantum problem is much richer due to the additional optimization over the measurement basis. A goodness of fit test for i.i.d quantum states is developed and a max-min characterization for the optimal measurement is introduced. We find the quantum measurement which leads both to the maximal Pitman and Bahadur efficiency, and determine the associated divergence rates. We discuss the relationship of the quantum goodness of fit test to the problem of estimating multiple parameters from a density matrix. These problems are found to be closely related and we show that the largest optimal error, determined by the smallest eigenvalue of the Fisher information matrix, is given by the divergence rate of the goodness of fit test.
\end{abstract}

\maketitle

\section{introduction}
The problem of quantum measurement has received a wide-ranging surge of interest because of  ground-breaking experiments in quantum information processing \cite{Aspect,Monroe,Bouwmeester,Furusawa,Pinkse,Riebe}. A fundamental feature of quantum measurements is the peculiar interplay between the quantum and classical world: a quantum measurement gives rise to "classical clicks", i.e. individual samples, and the only information that can be obtained when observing a quantum system is contained in the frequencies of the possible measurement outcomes. Let us consider a quantum experiment in which we receive a large but finite amount of identical copies of the state $\sigma$.  As the number of measurements that can be done is obviously bounded, there is no way by which two quantum systems whose density matrices are very close to each other can be distinguished exactly. In other words, it is fundamentally impossible to certify that a given system is in a particular quantum state $\sigma$: the only thing we can aim for is to certify that all the data collected in the experiment is compatible with the hypothesis that we sampled from the state $\sigma$.

Exactly the same problem is present in classical statistics \cite{Fisher}: It is impossible to certify that one is sampling from a given distribution, but one can only gain confidence that the samples are compatible or not with the fact that they are taken from a given distribution. Formally, the only thing achievable in a classical statistical experiment is to accept or reject a hypothesis. In the given setting, we take as the null hypothesis the fact that the distribution that we are sampling from has certain features, and we want to check whether the obtained data are compatible with this hypothesis. In practice, this means that a confidence interval has to be defined in which the hypothesis is accepted or rejected. The hypothesis is rejected when the experiment yields an outcome that was outside of this confidence interval, and accepted otherwise. Note that acceptance of the hypothesis does not imply that the hypothesis is true, it only indicates that the observed data are compatible with the hypothesis.

Such a  framework for hypothesis testing was developed one century ago by Pearson and Fisher \cite{Pearson,Fisher}, and forms the backbone for many more advanced techniques. One of the most successful tests is the so-called $\chi^2$ test. Its success has to do with the fact that it is universal \cite{Cramer}: The confidence intervals that can be defined are independent of the details of the distribution corresponding to the null hypothesis, as only the number of degrees of freedom plays a role. Also, the $\chi^2$ test is in practice already applicable when relatively few samples are taken. The $\chi^2$ test essentially measures the fluctuations around the expected frequencies of the possible outcomes: if those fluctuations are too small or too large, the hypothesis is rejected.

Fluctuations obviously also play a central role in quantum measurements. The expectation value of an observable is not something that can be measured, it can only be sampled, and we get an increasingly better precision the more measurements are being done. This actually means that the expectation value of an observable is not physical: only the individual samples (clicks) are physical. Expectation values can only be approximated by using the frequencies of the different outcomes. %As a consequence, the results of  quantum measurements should be reformulated in terms of observable quantities, i.e. the frequencies of clicks and not in terms of the expectation values which can never be observed directly. This actually means that the expectation value of an observable is not physical: only clicks are physical, and expectation values can be approximated using the frequencies of the different outcomes.

The topic of this paper is to make a detailed analysis of how the $\chi^2$ hypothesis test, when applied to the frequencies obtained from quantum measurements, reveals information about the underlying quantum states. A particular complication in the quantum setting that makes the problem much richer is the fact that we have the additional choice of the basis in which the measurements are done. Moreover, we will discuss a closely related question of optimally estimating multiple parameters in a given density matrix. We will find the measurement that minimizes the maximal quadratic error determined by the largest eigenvalue of the covariance matrix. The specific questions that we will address are:

\begin{enumerate}
\item How to set up the $\chi^2$ test in the quantum setting; how many degrees of freedom does the test have?
\item Suppose that we want to gain confidence that we prepared a certain quantum state $\sigma$ in the lab. What is the optimal POVM measurement such that, for all states for which $\|\rho-\sigma\|>\epsilon$, we would reject the hypothesis with the least amount of measurements if the state were $\rho$ instead of $\sigma$?
 \item What is the associated divergence rate for rejecting a false hypothesis?
\item What is the relationship between the resulting $\chi^2$ test and the quantum Fisher information used in parameter estimation?
\end{enumerate}

This paper fits into a long series of papers that were concerned with quantum parameter estimation and quantum hypothesis testing. A wealth of results has been reported in the seminal books of Helstrom \cite{Helstrom} and Holevo \cite{Holevo}, in a series of papers of Wootters \cite{Wootters} and other pioneers in the field of quantum information theory \cite{Braunstein,relent}. The more recent developments are covered in the books of Hayashi \cite{Hayashi} and Petz \cite{Petz}. Very recently, breakthroughs were obtained  in defining confidence intervals in the context of quantum tomography and testing of fidelity \cite{vanenk,Poulin,Liu,Christandl}. The present paper develops similar ideas in the framework of hypothesis testing. As opposed to Neyman-Pearson tests, the $\chi^2$ test is perfectly well defined without a need of formulating an alternative hypothesis. Such a situation arises precisely when we want to test whether a certain quantum state has been created in the lab.

In this context, a paper on quantum hypothesis testing for Gaussian states has recently been published \cite{Hayashi2} by Kumagai and Hayashi. The authors focus on the testing of families of quantum Gaussian states, which depend on two parameters, i.e. the number and the mean parameter. Depending on the particular parameter that is estimated, these tests can be seen as a quantum analogue of $\chi^2$-, t- and F- tests. The approach we take here is different in that we investigate the optimal application of the classical $\chi^2$-test for individual measurements and try to decide the best measurement strategy for arbitrary i.i.d states. Therefore, we will  focus on separable measurements, i.e. individual measurements on individual samples, as opposed to entangled measurements typically considered in Neyman-Pearson tests \cite{Audenaert,Skola}. The analysis presented here can therefore immediately be used in current experiments.\\

%\fixme{Put the reference and difference to the Hayashi paper here. We will also focus on separable measurements, i.e. individual measurements on individual samples, as opposed to entangled measurements such as typically considered in Neyman-Pearson tests \cite{Audenaert,Skola}. Therefore, the analysis presented here can immediately be used in current experiments.\\} %$\chi^2$ hypothesis testing is fundamentally different than the Neyman-Pearson test as usually discussed in quantum hypothesis testing \cite{Helstrom}. As opposed to Neyman-Pearson tests, the $\chi^2$ test is perfectly well defined without a need of formulating an alternative hypothesis. Such a situation arises precisely when we want to test whether a certain quantum state has been created in the lab. We will also focus on separable measurements, i.e. individual measurements on individual samples, as opposed to entangled measurements such as typically considered in Neyman-Pearson tests \cite{Audenaert,Skola}. 

Throughout this paper we will be working exclusively with states defined on finite dimensional Hilbert spaces $\cH \cong \bC^d$, which are isomorphic to the algebra of $d$-dimensional complex matrices $\cM_d \cong \bC^{d\times d}$. The states will be denoted by Greek letters $\rho,\sigma \in \cM_d$, with $\tr[\rho] = 1$ and $\rho \geq 0$. By means of the trace function we can introduce an inner product, often referred to as Hilbert-Schmidt scalar product, on the space of complex matrices  for $X, Y \in \cM_d$ via $\obraket{X}{Y} = \tr[X^\dag Y]$.  Together with this inner product $\cM_d$ can be regarded as a Hilbert space by itself and we can introduce a {\it bra-ket} notation for matrices $X \in \cM_d \cong \cc^{d\times d}$ via $\oket{X} = \sum_{ij} X_{ij} \ket{ij}$ and $\obra{X} = \sum_{ij} \overline{X}_{ij} \bra{ij}$ respectively. Moreover, we will be frequently making use of the identity $\oket{AXB} = A \otimes B^T \oket{X}$ when transitioning between the matrix and the vector representation of elements in $\cM_d$.

\section{Goodness of fit for quantum measurements}\label{goodFit}

We now come to the central part the paper, which is concerned with the problem of testing, whether the data acquired during an experiment is compatible with the fact that it originates from a given quantum state $\sigma$. Let us assume that we have an experimental quantum apparatus that supposedly spits out quantum states characterized by the density matrix $\sigma$. We would like to gain confidence that this hypothesis is true by performing measurements on it. We denote our {\emph hypothesis} $H$ by the fact that the $n$ samples we have obtained originate from doing quantum measurements on identical copies of the quantum state $\sigma$. The measurement will be described by a POVM with $r$ elements $\{E^i\}_{i=1\ldots r}$ which obey $\sum_i E^i = \openone$ and where all the individual elements are positive semi-definite $E^i \geq 0$. We say that the measurement has $r$ possible outcomes labeled by $i$ and associate a probability $p_i$ to each outcome which is given due to Born's rule by $p_i = \tr[E^i \sigma]$. If we record the number of times $n_i$ that we have obtained some outcome $i$, then we can construct the empirical distribution $f_i = n_i/n$ for the total number $n$ samples. By the law of large numbers \cite{Cramer}, we expect that as $n \rightarrow \infty$ the empirical distribution converges to $ f_i \rightarrow p_i$. 

%The most general measurement strategy would correspond to the case where different  positive operator valued measurements (POVM) $E^{\alpha,i}$ are chosen, with $E^{\alpha,i}\geq 0$, $\sum_{\alpha=1}^{r_i} E^{\alpha,i} = \openone$, and where the POVM $\{E^{.i}\}$ is measured a predetermined $k_i$ times (the fact that $k_i$ is not a random number is important for the determination of the number of degrees of freedom in the $\chi^2$ test). Such a typical setup for qubits would correspond to choosing $n_i=n/3$ with $n$ the total number of measurement done,  and von-Neumann measurements in the bases  $\sigma_x=E^{1,1}-E^{2,1}$, $\sigma_y=E^{1,2}-E^{2,2}$, and $\sigma_z=E^{1,3}-E^{2,3}$ respectively.  Alternatively, one could choose $k_1=n$ and do $n$ measurements with an informational complete POVM. We will henceforth consider the situation where only 1 POVM  is used to do the measurements; we will discuss how to modify the results in the case that different POVM are chosen deterministically, but the results essentially remain the same.

However, in any realistic scenario, we can only draw a finite number of samples. Due to the inherent randomness of the quantum measurements, there will be fluctuations. We will therefore have to consider a statistical test that does take these fluctuations into account. A test that is frequently considered in this scenario is the celebrated $\chi^2$ - test, originally introduced by Pearson \cite{Pearson}. The basis of this test is formed by the random variable
\be
	\chi^2 = \sum_{i=1}^r \frac{\left(n_i - n p_i \right)^2}{n p_i}.
\ee
This statistic is a good measure for testing whether we are sampling from $\{p_i\}$, as it measures the deviation of the empirical distribution $f_i$ from the ideal distribution $p_i$.
The $\chi^2$ - statistic is obviously a positive random variable. A crucial property of this random variable is the fact that its expectation value is independent of $n$ and is equal to the number of degrees of freedom, if the samples are indeed drawn from the distribution $\{p_i\}$ \cite{Aad}. If all the $r$ - probabilities $p_i$ are independent from each other, the total number of degrees of freedom is simply $r-1$. We will later consider tests, where the total number of $\{p_i\}$ can in fact vary, however, as these probabilities all stem from the same quantum mechanical state, the number of independent probabilities will always remain the same and is determined by the dimension of the Hilbert space.  In practice  statisticians use the following asymptotic form of the distribution for the $\chi^2$ variable:
\be\label{chi2distf}
	P_{r-1}(x) = \frac{1}{2^{\frac{r-1}{2}} \Gamma\left(\frac{r-1}{2}\right)} x^{\frac{r-3}{2}} \exp\left(-\frac{x}{2}\right).
\ee
For obvious reasons, this distribution is called the $\chi^2$-distribution, and is also the distribution which is obtained by summing up $r-1$ squares of random variables distributed following the normal distribution with expectation value $0$ and variance $1$ \cite{Aad}.  The power of the test $\alpha$ is obtained from choosing a threshold value $\chi^2_\alpha$ so that 
$ \alpha = \int_{\chi^2_\alpha}^\infty P_{r-1}(x) dx $. If the $\chi^2$-statistic grows larger than this threshold, the hypothesis $H$ is rejected.\\ 

Let us now study what will happen when the samples are not drawn from the quantum state $\sigma$ but from the state $\rho$. Then the measurement outcomes will not be distributed according to $p_i=\tr[E^i\sigma]$ but according to the distribution $q_i=\tr[E^i\rho]$. The expectation value of $\chi^2$ becomes

\begin{eqnarray}\label{chisquareMotivation}
\E_{q}[\chi^2] &=&\sum_i \frac{n^2 q_i^2+nq_i(1-q_i)}{np_i}-n \nonumber \\
&=&(n-1)\underbrace{\left(\sum_i \frac{\left(q_i-p_i\right)^2}{p_i}\right)}_{\chi^2(p,q)}+\left(\sum_i\frac{q_i}{p_i}\right)-1
\end{eqnarray}

The expectation value of $\chi^2$ grows linearly with the number of samples, and the multiplicative factor to this linear \emph{divergence} is defined as the $\chi^2$-divergence

\be 
\chi^2(p,q)=\sum_i \frac{\left(q_i-p_i\right)^2}{p_i}.
\ee

Obviously, if we would like to make the measurement which reveals the most information, it should be the one that would allow to reject the hypothesis $H$ as soon as possible if the hypothesis is false. That is, we want  the $\chi^2$ statistic to grow on average as fast as possible when we sample from a state different from $\sigma$. We therefore define an $\epsilon$-ball around our hypothesis state $\sigma$, and will optimize over all possible POVM measurements in such a way that we require that the (classical!) $\chi^2$ divergence with respect to all possible density matrices $\rho$ outside of this ball $\|\rho-\sigma\|\geq\epsilon$ is as large as possible. Due to the quadratic nature of the $\chi^2$ divergence, the natural norm to use is the Frobenius norm (i.e. $\|X\|=\sqrt{ \tr[X^\dagger X]}$); all bounds derived for the Frobenius norm, however, can be converted to any other norm such as the infinity or trace distance by using well known inequalities. The aforementioned discussion leads us to define the following quantity:

\begin{definition}\label{xi}
 The divergence rate $\xi$ for the quantum $\chi^2$ goodness of fit test for the state $\sigma$ is given by

 \be
 	\xi(\sigma) = \frac{1}{\epsilon^2}\max_{\{E^i\}} \min_{ \| \rho - \sigma \| \geq \epsilon} {\chi}^2(p,q),
 \ee

where we have defined the classical $\chi^2$-divergence

\be
{\chi}^2(p,q) = \left(\sum_i \frac{q_i^2}{ p_i  } - 1 \right),
\ee

with respect to the induced probability distributions $p_i = \tr[E^i \sigma]$ and  $q_i = \tr[E^i \rho]$. The optimization
is performed over all possible POVM $\{E^i\}_{i=1 \ldots r}$ and states $\rho$ for which $\|\rho-\sigma\| \geq\epsilon$ as measured by the Frobenius norm.
\end{definition}

Clearly, the optimal POVM should be an informationally complete POVM, as otherwise there would always be directions in which the divergence is zero. The properties of the optimal POVM will be discussed in the following section \ref{divRateAndoptPOVM}. Note that, due to the quadratic nature of $\chi^2$, $\xi(\sigma)$ is independent of $\epsilon$.  As will be proved in the next section, the divergence rate $\xi(\sigma)$ is guaranteed to lie in a small interval:

\be \frac{2}{3}\leq \xi(\sigma)\leq 1 \ee

This bound is actually very important: It shows that the prefactor of the linear term  of the expectation value of $\chi^2$ is independent of the dimension of the Hilbert space, which is of course crucial for the quantum $\chi^2$ hypothesis testing to make sense and to be scalable. Furthermore,  $\xi(\sigma)$ and the corresponding optimal POVM can be calculated exactly as the solution of a simple eigenvalue problem (see theorem \ref{theorem5}). As discussed later, the optimal POVM turns out to be optimal both in the sense of Pitman \cite{Pitman} and Bahadur \cite{Bahadur}.\\

A goodness of fit test protocol for the state $\sigma$ is then given as follows:

 \begin{enumerate}
	\item Choose the POVM $r$ element $\{{E^i}^*\}$ that optimizes $\xi$ as given in definition (\ref{xi}).
	
	\item Measure  $\{{E^i}^*\}$ on $n$ independent samples of the state $\rho$ and record the frequencies $n_i$ of the $i$'th
	      outcome.
	
	\item Compute the test statistic $c^2 = \sum_{i=1}^r \frac{\left(n_i - p_i n \right)^2}{p_i n}$, where
	      $p_i = \tr[E^i\sigma]$ corresponds to the hypothesis $H$.
	
	\item Reject the hypothesis with error probability $\alpha$ if $c^2 \geq \chi^2_\alpha$, where the constant $\chi^2_\alpha$
	      is determined via
	      \be
	      		\alpha = \int^{\infty}_{\chi^2_\alpha} P_{r-1}(x) dx
	      \ee
	
	\item If the test statistic $c^2$ is smaller than $\chi^2_\alpha$, we state that the observed data is consistent with the
	      hypothesis $H$ up to a statistical error $\alpha$. 

\end{enumerate}

Note that we assumed the large $n$ limit to compute the distribution function for the $\chi^2$ variable. This assumption is generally
well satisfied if we take sufficiently many samples.\\

If we now turn to the definition \ref{xi} of the divergence rate, we can give it  a meaningful interpretation in the light of the test protocol. The goal of the optimization is to construct a test, i.e. a quantum measurement, which rules out the hypothesis $H$ with as little samples as possible if is not true. That is, we want that the statistic $c^2$ grows as fast as possible with the number of samples $n$. In light of Eqn. (\ref{chisquareMotivation}), we see that the expectation value of the $\chi^2$ random variable grows linearly in the number of samples $n$ with the prefactor $\chi^2(p,q)$. In the case where $\rho = \sigma$ and thus $p=q$, i.e. the $H$ is true,  the classical $\chi^2$ vanishes and we obtain the expectation value $r-1$ and a standard deviation of $\sqrt{2(r-1)}$ \cite{Aad}. When $\rho\neq\sigma$, the goal is to find the measurement that reaches the critical region indicated by $\chi^2_\alpha$ as fast as  possible, in the worst case scenario.\\

We therefore have a class of estimators, parameterized by the different possible POVM $E$, and we want to find the most efficient one. Associated to every POVM $E$, there is a worst case state $\rho_E$ with $\|\rho_E-\sigma\|_2\geq \epsilon$ which gives rise to a divergence rate $\xi_E$. The expected number of samples $n$ needed to exceed the power $\alpha$ of the test statistic is given by the formula
 \be  (r-1)+(n-1)\epsilon^2\xi_E\simeq\chi^2_\alpha\ee
 or
 \be n\simeq\frac{\chi^2_\alpha-(r-1)}{\epsilon^2\xi_E} \ee
 
This is the number of expected samples which are necessary to reject the hypothesis if it is untrue.\\

Now there are several possible notions of efficiencies for asymptotic tests. For the so-called Pitman efficiency \cite{Pitman}, we compare tests in such a way that $\alpha$ is fixed but for which $\epsilon\rightarrow 0$ gradually, and look at the scaling of $n$ as a function of $\epsilon$.  Obviously, the POVM that minimizes $n$ is the one for which $\xi_E$ is maximal, i.e. the POVM that corresponds to the optimal one with respect to the definition  of $\xi(\sigma)$. Note that this POVM is also optimal according to Pitman for the maximum likelyhood. Different tests can also be compared with respect to the Bahadur efficiency \cite{Bahadur}. In the framework of Bahadur, $\epsilon$ is fixed, but the error $\alpha$ is made smaller and smaller (which corresponds to a larger and larger $\chi^2_\alpha$), and the scaling of $n$ with respect to $\alpha$ is compared. The optimal POVM which maximizes $\xi$  is obviously also the one with maximal Bahadur efficiency. The optimal quantum measurement is therefore the one with maximal Pitman and Bahadur efficiency within the class of all quantum $\chi^2$ tests.\\

Note that the standard deviation of $\chi^2$ is $\sqrt{2(r-1)}$. Therefore,  $\chi^2_\alpha-(r-1)$ for a fixed $\alpha$ but varying dimension of the Hilbert space is proportional to the square root of the number of degrees of freedom, i.e. linear in the dimension of the Hilbert space. 

\section{Divergence rate and optimal POVM}\label{divRateAndoptPOVM}

Let us next get some insights into the structure of the optimal POVM measurement. If the state $\sigma$ is full rank, the POVM must be informationally complete, so the number of POVM elements has to be at least equal to the square of the dimension of the Hilbert space, i.e. $r \geq d^2$, as otherwise there are always perturbations $X$ around  the state $\sigma$ for which $ \tr[E^iX]=0$. We will now prove that all the elements $E^i$ of the POVM must be pure, which is intuitively obvious. Then we will go on proving matching upper and lower bounds to the quantity $\xi(\sigma)$. The lower bound is constructive, and hence gives an explicit construction for the optimal measurement that maximizes the discriminating power.

\begin{lemma}\label{opt-POVM}  If the POVM $\{E^i\}$ is optimal in the sense that it maximizes the divergence rate, then all its elements can be chosen to be pure: $E^i=p_i\proj{\psi_i}$.
\end{lemma}

\emph{Proof:} Assume that the first element of the POVM with $r$ elements $\{E^i\}$ has rank $k_1>1$, i.e. $E^1=\sum_{l=1}^{k_1}p_l\proj{\psi_l}$. We will show that we can construct another POVM with $r+1$ elements which leads to a larger divergence rate, and for which the rank of $E^1$ is $k_1-1$ and the rank of $E^{r+1}$ is equal to $1$. Then the proof follows by induction. Let us therefore define $\tilde{E}^1=\sum_{l=2}^{k_1}p_l\proj{\psi_l}$ and $\tilde{E}^{r+1}=p_1\proj{\psi_1}$. We consider the change in the classical $\chi^2(p,q)$, which is then given by
\be
\sum_{i=1}^{r+1}\frac{(\tr{\tilde{E}^i(\rho - \sigma)})^2}{\tr{\tilde{E}^i \sigma}} - \sum_{i=1}^{r}\frac{(\tr{E^i(\rho - \sigma)})^2}{\tr{E^i \sigma}} \geq 0,
\ee
is positive for all possible density matrices $\rho$. We therefore need to show that the difference between the matrices
\be
\sum_{i=1}^{r+1} \frac{1}{\tr{\tilde{E}^i \sigma}} \oproj{\tilde{E}^i}  -  \sum_{i=1}^r \frac{1}{\tr{E^i \sigma}} \oproj{E^i} \geq 0,
\ee
is positive semi-definite. Since the old and new POVM coincide on almost all of the elements we are essentially left with the effectively two-dimensional matrix inequality 
\be
	\frac{1}{\tr{E^1 \sigma}}\oproj{E^1} \leq \frac{1}{\tr{\tilde{E}^1 \sigma}}\oproj{\tilde{E}^1} + \frac{1}{\tr{\tilde{E}^{r+1} \sigma}}\oproj{{E}^{r+1}},
\ee 
which can be verified easily when working in the basis $\oket{\tilde{E}^1}, \oket{\tilde{E}^{r+1}}$, since $\oket{E^1} = \oket{\tilde{E}^1} + p_1\oket{\tilde{E}^{r+1}}$. This immediately implies, 
that the new POVM has led to an increased divergence rate. Proceeding inductively, we are left with a POVM that consists only of rank-1 projectors. \qed

Note that in the proof we have modified the number of elements $r$ in the POVM. Moreover the optimization in definition \ref{xi}, does explicitly not specify the number of elements. 
Since we are considering a classical $\chi^2$ - goodness of fit test we need to consider the degrees of freedom \cite{Aad} of the test. Recall that the $\chi^2$-  distribution $P_r$  does neither depend on the original probabilities $\{p_i = \tr{E^i\sigma}\}$, nor on the total number of measurements, but only on the number of possible {\it independent} measurement outcomes $r-1$. This total number of degrees of freedom is equal to the number of independent $n_i$ that have to be specified. That is if the dimension of the  Hilbert space is $d$, we can have at most 
$r = d^2$ linearly independent POVM elements and thus we have to consider always a test with $d^2 -1$ degrees of freedom. For example, in the case of a POVM with 4 elements, $r=4$, but there is the constraint that $\sum_i n_i=n$, and we  have $3$ degrees of freedom. In the case of a single qubit, i.e. $d=2$, we have for the independent $\sigma_x,\sigma_y,\sigma_z$  measurements 6 frequencies $n_\alpha^i$, but only $3$ of them are independent, and hence we again have only $3$ degrees of freedom.\\

We are now ready to prove matching lower and upper bounds to $\xi(\sigma)$.

\subsection{Upper bound to the divergence rate}

An equivalent characterization of the divergence rate $\xi(\sigma)$ can be obtained by introducing the traceless operator $X=(\rho-\sigma)/\epsilon$:

\begin{equation}
\xi(\sigma)=\max_{\{E^i\}}\min_{X}  \obra{X}\left(\sum_i \oket{E^i} \frac{p_i}{\obraket{E^i}{\sigma}} \obra{E^i} \right) \oket{X}
\label{one}
\end{equation}

under the conditions:

\begin{eqnarray*}
&& E^i = |\psi^i\rangle\langle\psi^i|  \Sp \mbox{with} \Sp \langle\psi^i|\psi^i\rangle = 1  \Sp \mbox{and} \Sp \sum_i p_i E^i = \openone. \\
&& \tr[XX^\dagger] = 1  \Sp \mbox{with} \Sp  \tr[X]=0 \Sp \mbox{and} \Sp X=X^\dagger.
\end{eqnarray*}

The sum over $i$ is unlimited, i.e. there is no limit on the number of  POVM elements, and the dimension of $X$ is the dimension of the Hilbert space corresponding to $\sigma$, i.e. $d$-dimensional. Note that $\epsilon$ factored out due to the quadratic dependence on $\rho-\sigma=\epsilon X$. Without loss of generality, we will work in the basis in which $\sigma$ is diagonal:

\[\sigma=\sum_{\alpha=1}^d \lambda_\alpha | \alpha \rangle\langle \alpha |\]

with the eigenvalues $\lambda_\alpha=(s_\alpha)^2$ ordered in decreasing order. We will also assume that $\sigma$ is full rank; if this condition is not satisfied, then we can always perturb $\sigma$ infinitesimally, and take the limit at the end.

We will prove the following lemma:

\begin{lemma}
An upper bound to $\xi(\sigma)$ defined in (\ref{one}) is given by the smallest nonzero eigenvalue of the matrix

\be P_s\left(\sum_{\alpha=1}^d \frac{1}{1+\lambda_\alpha}\proj{\alpha}\right)P_s\ee

with $P_s$ the projector on the subspace orthogonal to the vector $\sum_\alpha \sqrt{\frac{\lambda_\alpha}{1+\lambda_\alpha}}\ket{\alpha}$.\\

A simple upper bound to this upper bound is
\[\xi(\sigma)\leq \frac{1}{1+\lambda_2}\leq 1\]
with $\lambda_2$ the second largest eigenvalue of $\sigma$.
\end{lemma}

Note that this upper bound lies between $2/3$ and $1$ for any density matrix.

\emph{Proof:} The proof of the theorem is a bit involved. In this proof, we will assume that the elements of the POVM are given by $p_i E^i$ with $E^i=\proj{\psi_i}$,
 $\braket{\psi_i}{\psi_i}=1$,  $\sum_i p_i E^i=\openone$ and $p_i \geq 0$, $\sum_i p_i =d$.

As a first step, we observe that as a consequence of the fact that $\sigma$ is diagonal we can twirl the POVM elements:

\[ \tr[E^i \sigma]=  \tr[E^i D(-\theta) \sigma D(\theta)]=\frac{\int d\theta_1d\theta_2\cdots  \tr[D(\theta)E^iD(-\theta) \sigma]}{\int d\theta_1d\theta_2\cdots }\]

 Here $D(\theta)$ is a diagonal matrix with elements $D_{kk}=\exp(i\theta_k)$. Therefore, two POVM related by $E^i=D(\theta)\tilde{E}^i D(-\theta)$ will give the same value in the optimization of (\ref{one}), as we can just transform the related $X$ to $\tilde{X}=D(-\theta)XD(\theta)$. It is therefore clear that an upper bound to (\ref{one}) is  obtained by solving the problem

\[\max_{\{E^i\}}\min_{X}  \frac{1}{(2\pi)^d} \int d\theta_1\cdots d\theta_d  \sum_i  \frac{p_i}{\obraket{E^i}{\sigma}} \left  | \obra{E^i}  D(\theta)\otimes D(-\theta)\oket{X} \right |^2  \]

as this forces one to use the same $X$ for different realizations of all equivalent POVM related by such a "gauge transformation". This is equivalent to saying that the minimum eigenvalue of a convex combination of operators with the same eigenvalues is always larger than the minimum of the individual eigenvalues. This twirling integration can be done exactly, and by using the cyclicity of the trace we get

\begin{eqnarray}
\hat{X} &= &\frac{1}{(2\pi)^d}\int d\theta_1\cdots d\theta_d   D(\theta)\otimes D(-\theta) \oproj{X} D(-\theta)\otimes D(\theta)   \no   &=&
\sum_{\alpha,\beta=1}^D X_{\alpha \alpha}X_{\beta \beta}| \alpha \rangle|\alpha\rangle\langle \beta|\langle \beta| +\sum_{\alpha \neq \beta} |X_{\alpha \beta}|^2 |\alpha\rangle\langle \alpha |\otimes |\beta\rangle\langle \beta| \nonumber
\end{eqnarray}

Substituting this into (\ref{one}), we get

\[\xi(\sigma)\leq \max_{E^i} \min_X \sum_{i}p_i\frac{\obra{E^i} \hat{X} \oket{E^i}}{\obraket{E^i}{\sigma}}\]
As $E^i=|\psi^i\rangle\langle\psi^i|$ are pure POVM elements,

\[\obra{E^i} \alpha \rangle\langle \alpha|\otimes |\beta \rangle\langle \beta\oket{E^i} = E_{\alpha \alpha}^i E_{\beta \beta}^i=\obra{E^i} \alpha\rangle|\alpha \rangle\langle \beta |\langle \beta \oket{E^i}.\]

Let's now define a new vector $|e^i\rangle$with $d$ components that contains the diagonal elements of $E^i$: $e^i_\alpha=E^i_{\alpha \alpha}$, and also the vector $|s\rangle$ with $d$ elements given by $s_\alpha=\sqrt{\lambda_\alpha}$ and $\lambda_\alpha$ the eigenvalues of $\sigma$.\\

Substituting all this into the previous expressions, we get

\[\xi(\sigma)\leq \max_{e^i}\min_X \sum_{i} p_i \frac{\sum_{\alpha \beta }\langle e^i|\alpha\rangle\langle \beta|e^i\rangle\left(|X_{\alpha \beta}|^2.(1-\delta_{\alpha \beta})+X_{\alpha \alpha}X_{\beta \beta}\right)}{\langle e^i|s^2\rangle}\]

Note that we have the constraints

\begin{eqnarray*}
\sum_i p_i \langle e^i|\alpha\rangle = 1 \Sp \mbox{and} \Sp \sum_\alpha X_{\alpha\alpha} = 0, \Sp \mbox{as well as} \Sp \sum_{\alpha \beta} |X_{\alpha \beta}|^2 = 1.
\end{eqnarray*}

The biggest problem in doing the optimization of equation (\ref{one}) is the presence of the denominator. Now is the time to get rid of it: we will choose $X$ such that

\[|X_{\alpha \beta}|^2.(1-\delta_{\alpha \beta})+X_{\alpha \alpha}X_{\beta \beta}=\langle \alpha|s^2\rangle\langle t^2|\beta\rangle + \langle \alpha |t^2\rangle\langle s^2| \beta \rangle=s_\alpha^2t_\beta^2+s_\beta^2t_\alpha^2\]

with the vector $|t^2\rangle$ with elements $\langle \alpha |t^2\rangle=|t_\alpha|^2$ still to be determined. Note that any choice of $X$ will give us an upper bound as long as the constraints above are satisfied. If it is possible to choose such a $|t\rangle$, then the upper bound becomes equal to

\begin{equation}
\xi(\sigma)\leq2\sum_i p_i \langle e_i|t^2\rangle\frac{\langle s^2|e^i \rangle}{\langle s^2|e^i \rangle}=2\sum_\alpha |t_\alpha|^2\label{two}
\end{equation}

This implies that such an $X$ and corresponding $t$ completely eliminates the $E^i$ from the upper bound, which was what we were looking for. It is indeed possible to choose such an $X$:

\begin{eqnarray*}
X_{\alpha \alpha}&=&\sqrt{2}s_\alpha t_\alpha\\
|X_{\alpha \beta}|^2&=&\left(s_\alpha t_\beta -s_\beta t_\alpha \right)^2
\end{eqnarray*}
The constraints on $X$ can now be written in  terms of the new variables $t_\alpha$:

\begin{eqnarray*}
0 &=& \sum_\alpha s_\alpha t_\alpha \\
1 &=& \sum_{\alpha \beta}|X_{\alpha \beta}|^2 = \sum_{\alpha \neq \beta} \left(s_\alpha t_\beta -s_\beta t_\alpha \right)^2+2\sum_\alpha \left(s_\alpha t_\alpha \right)^2\\
&=& 2\left(\sum_\alpha (1-s_\alpha^2)t_\alpha^2 -\sum_{\alpha \neq \beta} s_\alpha s_\beta t_\alpha t_\beta + \sum_\alpha  \left(s_\alpha t_\alpha \right)^2\right)\\
&=& 2\left(\sum_\alpha t_\alpha^2+\sum_\alpha s_\alpha^2 t_\alpha^2-\left(\sum_{\alpha} s_\alpha t_\alpha \right)^2\right) = 2\sum_\alpha (1+s_\alpha^2)t_\alpha^2
\end{eqnarray*}
Note that we made use of the normalization of $\sigma$ in the form of $\sum_\alpha s_\alpha^2 = 1$ and also of the constraint $\sum_\alpha s_\alpha t_\alpha = 0$. Rescaling $t^2$ by a factor of $2$, we get  the optimization problem:

\begin{eqnarray*}
{\rm minimize} & \sum_{\alpha =1}^d t_\alpha^2\\
{\rm under \hspace{.1cm} the \hspace{.1cm} condition} & \sum_{\alpha =1}^d s_\alpha t_\alpha=0\\
{\rm and} & \sum_{\alpha=1}^d (1+s_\alpha^2)t_\alpha^2=1
\end{eqnarray*}

This optimization problem can actually be written as an eigenvalue problem: define $y_\alpha=\sqrt{1+s_\alpha^2}t_\alpha$ and $P_s$ the projector on the space orthogonal to the vector with components $s_\alpha/\sqrt{1+s_\alpha^2}$.  Then the upper bound is given by the second smallest eigenvalue (the smallest being 0) of the matrix

\be P_s\sum_\alpha \frac{1}{1+s_\alpha^2}|\alpha \rangle\langle \alpha| P_s. \label{uppp} \ee

This is the upper bound that we set out to prove. A simple upper bound to this upper bound can be found. By making use of the interlacing properties of eigenvalues of submatrices, we therefore know that the eigenvalues of this matrix obey

\[  \mu_1=0 \leq \frac{1}{1+s_1^2}\leq \mu_2 \leq \frac{1}{1+s_2^2} \leq ...\]

which proves that
\[\xi(\sigma)\leq \frac{1}{1+s_2^2}\leq 1.\]

This concludes the proof.  \qed

\subsection{Lower bound to the divergence rate}
Let us next prove a lower bound to the divergence rate $\xi(\sigma)$. For this, we will have to guess a class of good POVM. We will do the optimization over the class of POVM parameterized by a single parameter $0\leq p\leq 1$:

\begin{eqnarray}
1\leq i\leq d: & E^i &= (1-p) \proj{i}\\
j> d: & E^j & =  c(p)\proj{\chi_j}\\
& |\chi_j\rangle &= \frac{1}{\sqrt{d}}\sum_k e^{i\theta^k_j}|k\rangle,
\end{eqnarray}
where the $\{\ket{i}\}$ label the eigenstates of $\sigma$. All $\ket{\chi_j}$ are chosen such that they have the same overlap with $\sigma$: $\bra{\chi_j}\sigma\ket{\chi_j}=1$. Those states $\ket{\chi_j}$ are hence only susceptible to the off-diagonal elements of $\sigma$. In the case of $d$ a prime or a power of prime, a possible choice of such a basis is given by the mutually unbiased basis, but as we only require unbiasedness with the standard basis, such a basis can easily be constructed in any dimension, e.g. by choosing basis labeled by the angles $\{\theta_{j}^k\}$. We will choose such a basis that is  invariant under any similarity transformation with diagonal elements $D_{kk}=\exp(i\theta_k)$ (which is always possible), such that we have $\sum_{j>d}E^j=c(p)\sum_{j>d}\proj{\chi_j}=p\openone$. This defines $c(p)$ which we do not have to determine explicitely. It follows that

\begin{eqnarray}
\sum_{j>d} \oket{E^j}\frac{1}{ \obraket{E^i}{\sigma}}\obra{E^j} &=& \frac{c(p)}{1/d}\sum_{j>d}\ket{\chi_j}\ket{\overline{\chi}_j}\bra{\chi_j}\bra{\overline{\chi}_j} \\ \nonumber 
&=&p\left(\sum_{i\neq j}\proj{ij}+\sum_{i,j}\ket{ii}\bra{jj}\right).
\end{eqnarray}
This follows from the fact  that the operator is invariant under twirling, and also because

\[ p.d= \tr \sum_j E^j=c(p) \tr \sum_j \proj{\chi_j}=c(p) \tr \sum_j \proj{\chi_j}\otimes \proj{\overline{\chi}_j}.\]

With those choices, there is only one parameter left, i.e. the weight $p$ that weights the diagonal versus the off-diagonal parts of the density matrix $\sigma$. A lower bound on $\xi(\sigma)$ can now be obtained by the following optimization:

\be \max_p\min_X\obra{X} \underbrace{(1-p)\sum_{i=1}^d \frac{1}{\lambda_i}\proj{ii}+p\left(\sum_{i\neq j}\proj{ij}+\sum_{i,j}\ket{ii}\bra{jj}\right)}_{Q}\oket{X} \ee

with $X=(\rho-\sigma)/\epsilon$  a traceless Hermitean operator with norm $\|X\|_2=1$. We therefore want to make the smallest eigenvalue of the matrix $Q$ as large as possible, as this eigenvalue provides a lower bound to $\xi$. The matrix $Q$ is a direct sum $Q_1\oplus Q_2$ where $Q_1$ is $p$ times the identity matrix on the subspace spanned by $\ket{ij},i\neq j$, and $Q_2$ the $d\times d$ matrix

\be Q_2=(1-p)\sum_{\alpha=1}^d \frac{1}{\lambda_\alpha}\proj{\alpha} + p\sum_{\alpha,\beta=1}^d \ket{\alpha}\bra{\beta} \ee

where we identified $\ket{\alpha}=\ket{ii}$. Actually, this is not entirely correct, as we still have to include the constraint that $ \tr[X]=0$. This can easily be incorporated by projecting $Q_2$ on  the subspace orthogonal to $\ket{\Omega}=1/\sqrt{d}\sum_\alpha\ket{\alpha}$.  Given $P=\openone-\ket{\Omega}\bra{\Omega}$, we therefore define $\tilde{Q}_2=PQ_2P$.\\

The smallest eigenvalue of $Q_1$ is obviously proportional to $p$, while the smallest eigenvalue of $\tilde{Q}_2$ is monotonically decreasing with $p$.  Therefore, the optimal value of $p$ will be the one for which the smallest eigenvalues of $Q_1$ and $\tilde{Q}_2$ coincide. This is equivalent to determining the largest $p$ for which

\be (1-p)\sum_{\alpha=1}^d \frac{1}{\lambda_\alpha}P\proj{\alpha}P + p\sum_{\alpha,\beta=1}^d P\ket{\alpha}\bra{\beta}P \geq p P \ee

which is in turn equivalent to maximizing $p$ such that
\be \sum_\alpha \frac{1}{\lambda_\alpha}P\proj{\alpha}P\geq p \left(\sum_\alpha \frac{1}{\lambda_\alpha}P\proj{\alpha}P-\sum_{\alpha \neq \beta}P\ket{\alpha}\bra{\beta}P\right). \ee

This optimal $p$, which is the lower bound we were looking for, is then given by

\be p=\frac{1}{\mu(S)} \ee
with $\mu$ the largest eigenvalue of the matrix

\be S=\openone-\frac{1}{d}\sum_{\alpha,\beta}\ket{\alpha}\bra{\beta}+\sum_\alpha \lambda_\alpha\proj{\alpha}-\sum_{\alpha,\beta}\lambda_\alpha \lambda_\beta\ket{\alpha}\bra{\beta} \ee

which is equivalent to $1$ plus the largest eigenvalue of the pseudo-inverse of the matrix $P\sigma^{-1} P$:

\be \tilde{S}=\sum_\alpha \lambda_\alpha\proj{\alpha}-\sum_{\alpha,\beta}\lambda_\alpha \lambda_\beta\ket{\alpha}\bra{\beta} \ee

$\tilde{S}$ is again the  generator of a semi-group, and hence all its eigenvalues are larger or equal to zero. It is equal to zero for pure states, and the maximal possible eigenvalue is equal to $1/2$ and is obtained for the case $\lambda_1=\lambda_2=1/2$, $\lambda_{i>2}=0$.  Those 2 cases correspond to $\xi=1$ and $\xi=2/3$ respectively. It can easily be shown that the pseudo-inverse of the matrix $S$ has the same eigenvalues as the matrix (\ref{uppp}). This means that our lower bound
coincides with the upper bound! We have therefore proven:

\begin{theorem}\label{theorem5}
The divergence rate $\xi(\sigma)$ is equal to $\xi(\sigma) =  1/(1+\mu(S))$ with $\mu(S)$ the largest eigenvalue of the matrix

\be S=\sum_\alpha \lambda_\alpha\proj{\alpha}-\sum_{\alpha,\beta}\lambda_\alpha \lambda_\beta\ket{\alpha}\bra{\beta} \ee

where ${\lambda_\alpha}$  are the eigenvalues of $\sigma$ and $\ket{\alpha}$ the corresponding eigenvectors. In particular, this implies that

\be \frac{2}{3}\leq \xi(\sigma)\leq 1 \ee

with the value of $2/3$ obtained in the case where $\lambda_1=\lambda_2=1/2$, $\lambda_{i>2}=0$, and the value $1$ when $\sigma$ is a pure state.
A possible choice for a POVM that gives the optimal error rate is given as follows:

\begin{eqnarray}\label{optPOVM}
1\leq i\leq d: & E^i &= (1-\xi) \proj{i}\\
j> d: & E^j & =  c(\xi)\proj{\chi_j}\\
& |\chi_j\rangle &= \frac{1}{\sqrt{d}}\sum_k e^{i\theta^k_j}|k\rangle
\end{eqnarray}

with $c(\xi)$ and the angles $\{\theta^k_j\}$ chosen such that the POVM is informationally complete and that

\[ \sum_{j>d} E^j  =  \xi\openone \]

\end{theorem}

Note that the degrees of freedom in the $\chi^2$ distribution corresponding to this optimal POVM can easily be reduced by dividing the POVM up in several resolutions of the identity, and fixing the number of times those different measurements are done by a fraction corresponding to their weight given in the theorem. For example, let us assume that the $\ket{\chi_j}$ can be divided up into $d$ orthonormal basis (as e.g. in the case of mutually unbiased bases), and that we want to do a total of $N$ measurements. The von Neumann measurement in the basis $\ket{i}$ can then be done $(1-\xi).N $ times and the other von-Neumann measurements $\xi/d.N$ times. The total number of degrees of freedom for the corresponding $\chi^2$ distribution is then given by $(d+1).d-(d+1)=d^2-1$ which is indeed equal to the total number of degrees of freedom in the density matrix. It is clear that exactly the same arguments for the error exponent carry through in this case.

\subsection{Examples of divergence rates}
Let us next look at some specific examples. A special role is played by the second largest eigenvalue of $\sigma$: $\xi$ is minimized when $\lambda_2$ is maximal, and maximized when $\lambda_2$ is minimal. The maximal divergence rate is obviously obtained for pure states and is exactly given by $1$:

\be \xi\left(\proj{\psi}\right)=1 \ee

Furthermore, the states for which it is most difficult to do hypothesis testing are the ones corresponding to projectors on a 2-dimensional subspace:

\be \xi\left(\frac{1}{2}\sum_{\alpha=1}^2\proj{\alpha}\right)=\frac{2}{3} \ee

$\sigma=P/2$. However, there is clearly not a big discrepancy between $2/3$ and $1$, so  the test will perform well for any state $\sigma$.

Another interesting class of states contains all maximally mixed states: Here

\be \xi\left(\openone/d\right)=\frac{1}{1+1/d}.\ee

Finally,  $\xi(\sigma)$ can be calculated analytically for any density matrix defined on a 2-level system:

\be \xi(\sigma) = \frac{1}{1+2\lambda_1\lambda_2} \ee

Following the constructive proof of the lower bound, A $POVM$ with $6$ elements that saturates this  is given by

\begin{eqnarray}
 \begin{array}{lll}E^1= (1-\xi(\sigma)) \proj{0} ,& E^3=  \frac{\xi(\sigma)}{2}\proj{+},& E^5=  \frac{\xi(\sigma)}{2}\proj{i}, \vspace{0.3cm} \\  \nonumber 
 E^2= (1-\xi(\sigma)) \proj{1},& E^4= \frac{\xi(\sigma)}{2} \proj{-},&   E^6 = \frac{\xi(\sigma)}{2}\proj{-i}   \end{array}
\end{eqnarray}

where we work in the basis where $\sigma$ is diagonal, and  with $\ket{\pm}$ and $\ket{\pm i}$ the eigenbasis of the Pauli matrices $\sigma_x$ and $\sigma_y$. An optimal $\chi^2$ test with $3$ degrees of freedom on $N$ samples is then obtained by doing $(1-\xi).N$ measurements in the computational basis and $\xi.N/2$ in both the $\sigma_x$ and $\sigma_y$ basis.

\section{The $\chi^2$ - test and its connection to quantum parameter estimation}

We will now turn to a closely related topic, the estimation of parameters for a family of density matrices $\sigma$, that depends on a set of parameters $\underline{\theta} = (\theta_1,\ldots,\theta_m) \in \cR^m$. First, we discuss the well known scenario of estimating a single parameter and relate this to a family of quantum $\chi^2$ divergences. We will then turn to a measurement strategy for estimating multiple parameters from a quantum distribution. We will discuss a novel optimality criterium for the best measurement strategy in the case of multiple parameters, which aims at minimizing the largest eigenvalue of the classical Fisher information matrix. It turns out, that this approach is strongly related to the $\chi^2$ hypothesis test discussed in the previous section, when we seek to estimate all the parameters of the density matrix $\sigma$.

It is a well known result in classical estimation theory, that the covariance matrix of a set of unbiased, sufficient statistics defined as  $\hat{\underline{\theta}} : x \in X \raw (\hat{\theta}_1(x),\ldots,\hat{\theta}_m(x)) \in \cR^m$ is lower bounded by the Fisher information matrix \cite{Aad}, when we consider samples $x \in X$ from some sufficiently smooth distribution $p_k(\underline{\theta})$. That is, one always obtains the bound
\bq
	Cov(\hat{\underline{\theta}},\hat{\underline{\theta}}) \geq J^{-1}({\underline{\theta}}),
\eq
as a semidefinite inequality for the covariance matrix
\be
\left[Cov(\hat{\underline{\theta}},\hat{\underline{\theta}})\right]_{ij} = \E_{p(\underline{\theta})}\left[(\hat{\theta}_i - \theta_i)(\hat{\theta}_j - \theta_j)\right]
\ee
and the Fisher information matrix 
\be
\left[J({\underline{\theta}})\right]_{ij} = \sum_k  \left( \frac{\partial}{\partial \theta_i}\log p_k(\underline{\theta}) \right)\left( \frac{\partial}{\partial \theta_j}\log p_k(\underline{\theta}) \right) p_k(\underline{\theta}) . 
\ee
Moreover, it known that in the asymptotic limit, the lower bound can actually be met by a maximum likelihood estimator \cite{Aad}.\\

Let us first consider the case, where $p_\theta$ depends only on a single parameter $\theta \in \cR$.  In this case it is easy to see that the Fisher information is equal to 
\be J(\theta) = \lim_{\epsilon \raw 0} \frac{1}{\epsilon^{2}}\chi^2(p_\theta + \epsilon \frac{\partial}{\partial \theta}p_\theta,p_\theta). \ee 

The problem of estimating a single parameter of the family of density matrices $\sigma(\theta)$, with $p_k(\theta) = \tr{[E^k \sigma(\theta)]}$, was first considered by Holevo \cite{Holevo}. Later an alternative proof of the quantum Fisher information lower bound was given by Braunstein and Caves in the context of the statistical geometry of quantum states \cite{Braunstein}. The authors took a different approach from that in \cite{Holevo}, by first giving the general lower bound to the variance of any unbiased estimator in terms of the classical Fisher information and then, optimizing the Fisher information over all possible POVM $\{E_k\}$ to obtain a lower bound in terms of the quantum Fisher information. It is shown, that this lower bound can in fact be obtained for a particular measurement. For  the general case, the attainability of these bounds has  been discussed in \cite{Gill,Luati}. We will now establish a connection to  a family of quantum versions of the $\chi^2(\rho,\sigma)$-divergence, introduced in the paper \cite{chisquared},  to study the convergence and relaxation rates \cite{Reeb} of completely positive maps and general dissipative quantum systems. All members of this class of quantum $\chi^2$-divergences reduce to the classical $\chi^2$ divergence when $\rho$ and $\sigma$ commute. The proof presented here has no similarity to this original proof; a central role is played by Woodburry's matrix identity \cite{WoodBerry}.\\

The formulation of the quantum versions of the  $\chi^2$-divergence follows from the framework of monotone Riemannian metrics  \cite{Morozova,Petz0,Petz1,Petz2,Petz3,LR} and can be seen as a special case of this family of metrics. It follows from the analysis of monotone Riemannian metrics that the family of $\chi^2$-divergences has a partial order with a smallest and largest element.  A special role was played by the Bures $\chi^2$ divergence \cite{Bures,Uhlman}, as it is always the smallest one of those quantum divergencies. It is defined as

\be \chi^2_B(\sigma,\rho)=\tr\left[(\rho-\sigma)\Omega_\sigma(\rho-\sigma)\right]
\ee

with $\Omega_\sigma$ the superoperator whose inverse is given by

\be \Omega_\sigma^{-1}(X)=\frac{\sigma X+X\sigma}{2} \ee

Let us now show that an operational meaning can be given to this quantity by comparing it to the  classical $\chi^2$ divergence maximized over all possible quantum measurements.

\begin{lemma}\label{chi2-dev}
For two states $\sigma $ and $\rho$ we denote the probability distributions  $p_i = \tr[E^i\sigma]$ and $q_i  = \tr[E^i \rho]$
for some POVM $\{E^i\}_{i= 1 \ldots r}$. Then, the Bures $\chi^2_B$-divergence is equal to the maximum value of $\chi^2(p,q)$ when optimized over all possible POVM measurements:

\be\label{max-chi2-dev}
	\chi^2_B(\rho,\sigma) = \max_{\{E^i\}} {\chi}^2(p,q),
\ee

Furthermore, the measurement maximizing this $\chi^2$ divergence is a projective von Neumann measurement in the eigenbasis of $\Omega_\sigma(\rho) = \sum_i \lambda_i \proj{\psi_i}$.
\end{lemma}

\proof{
Let us first prove that Bures $\chi^2$ divergence forms an upper bound to the $\chi^2$ divergence with respect to any POVM $\{E^i\}$.
Let us denote by $\hat{\Omega}_\sigma^B$ the matrix representation of $\Omega_\sigma^B$ on the vector space $\cc^{d \times d}$. 
We then have that
 \be 
 \hat{\Omega}_\sigma^B = \frac{2}{ \sigma \otimes \openone + \openone \otimes \sigma^T}. 
 \ee
It is easy to see that the $\chi^2$ divergence  is given by

\be \chi^2(p,q)=\obra{\rho}\sum_i\frac{\oproj{E^i}}{\obraket{E^i}{\sigma}}\oket{\rho} -1\ee

and the Bures divergence by

\be \chi^2_B(\sigma,\rho)=\obra{\rho} \frac{2}{ \sigma \otimes \openone + \openone \otimes \sigma^T} \oket{\rho} -1 \ee

It is therefore enough to prove the semidefinite matrix inequality

\be \frac{2}{ \sigma \otimes \openone + \openone \otimes \sigma^T}-\sum_i\frac{\oproj{E^i}}{\obraket{E^i}{\sigma}}\geq 0 \label{ineq}\ee

holds for all possible POVM $\{E^i\}$. A matrix is positive if and only if its inverse is positive, and the inverse can easily be calculated by making use of Woodburry's Identity \cite{WoodBerry}

 \be\label{WoodBerry}
 	(A - U C U^\dagger)^{-1} = A^{-1} +  A^{-1}U\left( C ^{-1} - U^\dagger A^{-1} U \right)^{-1} U^\dagger A^{-1}.
 \ee

Equation (\ref{ineq}) is exactly of that form by choosing an orthonormal basis $|i\rangle$ with a number of elements equal to the total number of POVM elements and

\begin{eqnarray*}
A =  \frac{2}{ \sigma \otimes \openone + \openone \otimes \sigma^T} \Sp \mbox{and} \Sp U  = \sum_i \oket{E^i}\bra{i}, \; \mbox{as well as } \; C = \sum_i \frac{ \proj{i}}{\tr\left[E^i \sigma \right]}.
\end{eqnarray*}

As the matrix $A$ is obviously positive, (\ref{ineq}) will hold if

\begin{eqnarray} 
&&C^{-1}-U^\dagger A^{-1} U = \\ \nonumber 
&& \sum_i \tr[E^i\sigma]\proj{i}-\sum_{ij}\ket{i}\bra{j}\obra{E^i}\frac{\sigma\otimes\openone+\openone\otimes\sigma^T}{2}\oket{E^j}\geq 0.
\end{eqnarray}

or equivalently if the matrix $L = \sum L_{ij} \ket{i}\bra{j}$, with the entries
\begin{eqnarray}
	L_{ij} =\left \{ \begin{array}{l l}   \tr\left[E^i(\openone - E^i)\sigma\right]  & \Sp i = j  \\\nonumber
	                                                  -\frac{1}{2}\left(\tr\left[E^iE^j\sigma\right] + \tr\left[E^jE^i\sigma\right] \right) & \Sp  i \neq j,  \end{array} \right.
\end{eqnarray}\vspace{0.25cm}

is positive semi-definite. Recall that we have shown, c.f. lemma \ref{opt-POVM}, that the measurement for which the maximum in (\ref{max-chi2-dev}) is obtained are pure and of the form $E^i = p_i \proj{\psi_i} = \proj{\tilde{\psi}_i}$. We therefore have that 
\be
	L = \frac{1}{2} \left(S +  S^T\right), 
\ee
where $T$ denote the transpose in the basis spanned by $\{\ket{i}\}$. The components of the matrix $S$ are  given by
\be
	S_{ij} = \bra{\tilde{\psi}_i}\sigma\ket{\tilde{\psi}_j}\left( \delta_{ij} - \braket{\tilde{\psi}_j}{\tilde{\psi}_i}\right).
\ee
Note, that the matrix $ [\bra{\tilde{\psi}_i}\sigma\ket{\tilde{\psi}_j}]_{ij}$ is always positive semi-definite since $\sigma$ is.  Moreover, we have that 
the matrix $\left( \delta_{ij} - \braket{\tilde{\psi}_j}{\tilde{\psi}_i}\right)$ is also always positive, since the $\ket{\tilde{\psi}_i}$ span the rows of an isometry. 
We therefore have that $S$ is the Hadamard product of two positive semi-definite matrices and is therefore positive semi-definite itself. Hence, $L$ is positive semi-definite ad the inequality (\ref{ineq}) follows. \\

Note that we can make $L$ equal to zero by choosing all $POVM$ elements orthogonal to each other, i.e. by choosing a von Neumann measurement.  The null space of the matrix occurring in (\ref{ineq}) can now easily be seen to be spanned by the vectors in $A^{-1}U$. $\rho$ will therefore be in the null space and saturate the inequality iff there exist numbers $\{\lambda_i\}$ for which

\be \oket{\rho}=\sum_i \lambda_i \bra{i}A^{-1}U\ket{i}=\sum_i \lambda_i \frac{\sigma\otimes\openone+\openone\otimes\sigma^T}{2}\oket{E^i}.\ee

By writing $E^i=\proj{\psi_i}$, this equation is equivalent to

\be\label{GoodMeasure} \sum_i\lambda_i \proj{\psi_i}=\Omega_{\sigma}^B(\rho) \ee

which shows that a von Neumann measurement in the eigenbasis of $\Omega_{\sigma}^B(\rho)$ will give equality.
\qed}

From this, the result of Braunstein and Caves is now obtained by defining  the quantum Fisher information as
\be
J_{QM}(\theta) = \tr{\left[\frac{\partial}{\partial\theta}\sigma(\theta) \Omega^B_\sigma\left(\frac{\partial}{\partial\theta}\sigma(\theta)\right)\right]} = \lim_{\epsilon \raw 0} \frac{1}{\epsilon^2} \chi^2_B\left(\sigma + \epsilon\frac{\partial}{\partial\theta}  \sigma(\theta) ,\sigma\right).
\ee
This shows that the quantum $\chi^2$ divergences have indeed an operational meaning.\\

Let us now turn to the problem of estimating multiple parameters $(\theta_1,\ldots,\theta_m)$. In analogy to the classical Fisher information matrix one commonly proceeds \cite{Helstrom,Holevo} to define the quantum Fisher information matrix by
\be
\left[ J_{QM}(\underline{\theta})	\right]_{ij} =  \tr{\left[\frac{\partial}{\partial{\theta_i}}\sigma(\underline{\theta}) \Omega^B_\sigma\left(\frac{\partial}{\partial{\theta_j}}\sigma(\underline{\theta})\right)\right]}.
\ee

Indeed this matrix does form an upper bound to  the classical Fisher information matrix for all POVM. That is we always have that $J(\underline{\theta}) < J_{QM}(\underline{\theta})$  and by this we have a chain of inequalities for the covariance matrix
\be\label{naughtyBound}
	Cov(\hat{\underline{\theta}},\hat{\underline{\theta}}) \geq J^{-1}({\underline{\theta}}) > J^{-1}_{QM}(\underline{\theta}).
\ee
Recall, that for a single parameter the bound by the classical Fisher information matrix could be saturated by an appropriately chosen POVM, (\ref{GoodMeasure}). This is no longer the case for the the estimation of multiple parameters \cite{Gill,Luati}. That is, the final lower bound in (\ref{naughtyBound}) can not be obtained. Let us therefore propose an alternative optimality criterium for the estimation of multiple parameters. Rather than attempting to saturate the matrix inequality (\ref{naughtyBound}), we will try to minimize the largest possible error, i.e. the largest eigenvalue of the inverse classical Fisher information matrix. That is, we consider the optimization problem
\be
	\min_{E^i} \lambda_{max} (Cov(\hat{\underline{\theta}},\hat{\underline{\theta}})) \geq  \min_{E^i}  \min_{\|x\|_2 = 1} \bra{x}J^{-1}({\underline{\theta}}) \ket{x}.
\ee   
Equivalently, we want to maximize the smallest eigenvalue of $J({\underline{\theta}})$.\\ 

We consider the following set up. Assume we have an orthonormal Hermitian,matrix basis $\{F_\alpha, \I d^{-1}\}_{\alpha = 1\ldots d^2-1}$ of the space $\cM_d$ with respect to the canonical Hilbert-Schmidt product. We then consider estimating $d^2-1$ parameters $\theta_\alpha$ of the density matrix $\sigma(\underline{\theta}) = \frac{1}{d}\I + \sum_{\alpha} \theta_\alpha F_\alpha$. In light of the previous discussion we state the following theorem:

\begin{theorem}\label{lastTheorem}
Let  $\sigma(\underline{\theta}) = \frac{1}{d}\I + \sum_{\alpha} \theta_\alpha F_\alpha$, where the $\{\theta_\alpha\} \subset \cR^{d^2 -1}$ are chosen so that $\sigma(\underline{\theta}) > 0$.
The maximum over all POVM  $\{E^i\}$ of smallest eigenvalue of the classical Fisher information matrix $J(\underline{\theta})$ with $p_i(\underline{\theta}) = \tr{[E^i \sigma(\underline{\theta})]}$ is given by
\be\label{FisherEig}
	\max_{E^i}  \min_{\|x\|_2 = 1} \bra{x} J(\underline{\theta}) \ket{x} = \xi(\sigma),
\ee
where $\xi(\sigma)$ is the divergence rate determined by  theorem \ref{theorem5}. The optimal POVM that saturates the bound is given by  Eqn.  (\ref{optPOVM}). 

Moreover, let $Cov(\underline{\hat{\theta}},\underline{\hat{\theta}})$ denote the covariance matrix of an unbiased statistic for $\underline{\theta}$, then the operator norm is bounded by
\be\label{CovBound}
	\|Cov(\underline{\hat{\theta}},\underline{\hat{\theta}})\|_{2 \raw 2} \geq \frac{1}{\xi(\sigma)}. 
\ee
This lower bound can be obtained in the asymptotic limit.
\end{theorem}

\proof{With $p_i(\underline{\theta}) = \tr{[E^i\sigma(\underline{\theta})]}$  and the particular form of $\sigma$, we immediately have that $\frac{\partial}{\partial\theta_\alpha} p_i(\underline{\theta}) = \tr{[E^i F_\alpha]}$. Let us now consider 
\be
	\bra{x} J(\underline{\theta}) \ket{x} = \sum_i \sum_{\alpha \beta} x_\alpha x_\beta \frac{\tr{[E^i F_\alpha]}\tr{[E^i F_\beta]}}{\tr{[E^i\sigma]}} =  \sum_i  \frac{\tr{[E^i X]^2}}{\tr{[E^i\sigma]}},
\ee
with $X = \sum_\alpha x_\alpha F_\alpha$. Note that since the $F_\alpha$  are Hermitian and traceless, we have that $\tr{X} = 0$ and $X^\dagger = X$. Due to the normalization of $\ket{x}$,
and the orthonormality of $F_\alpha$ we also have that $\tr[{X^\dagger X}] = 1$. The problem in eqn. (\ref{FisherEig}) is therefore identical to the previous optimization problem (\ref{one}). The solution to the problem is given in theorem \ref{theorem5}. The final inequality (\ref{CovBound}) is a direct consequence of the fact that $J^{-1}(\underline{\theta})$ is a lower bound for the covariance matrix \cite{Aad} attained in the asymptotic limit. \qed }

\vspace{0.5cm}
Although the optimization problem considered in this theorem \ref{lastTheorem} is in fact identical to that of theorem \ref{theorem5}, the interpretations are quite different. In the original problem (theorem \ref{theorem5}), we were looking for the best measurement that would allow us to reject the false hypothesis $H$ as soon as possible. Here, we determined the largest possible error for estimating multiple parameters and tried to find a POVM that makes this error as small as possible. \\

Let us now discuss the relationship to the canonical  quantum Fisher information matrix bound. In particular, let us compare the resulting operator norms of both approaches. That is, we define the smallest eigenvalue of $J_{QM}$ as
\be
	\tilde{\xi}(\sigma) = \min_{\|x\|_2 = 1} \bra{x}J_{QM}(\underline{\theta})\ket{x}.
\ee
We again focus on $\sigma(\underline{\theta}) = \frac{\I}{d} + \sum_\alpha \theta_\alpha F_\alpha$ as we have done previously. This optimization problem is equivalent to 
\be
	\tilde{\xi}(\sigma) = \min_{\tr{X^\dagger X} = 1} \obra{X} \frac{2}{\sigma\otimes\I + \I \otimes \sigma^T}\oket{X},
\ee   
for $X \in M_d$ Hermitian and traceless.  Working in the eigenbasis of $\sigma = \sum_i \lambda_i \proj{i}$. with $\lambda_0 \geq \lambda_1 \ldots \geq \lambda_d$, we see that the operator 
$(\frac{1}{2}(\sigma \otimes \I + \I \otimes \sigma^T ))^{-1}$ is diagonal in the matrix units $\ket{ij}$. If we take into account that we optimize over $X$ with the aforementioned constraints, we obtain easily that
\be
	\tilde{\xi}(\sigma) = \frac{2}{\lambda_0 + \lambda_1}.
\ee 

The important point is that $\xi(\sigma)$ is always a better bound on the operator norm of the covariance matrix than the largest eigenvalue $\tilde{\xi}(\sigma)$ of the quantum Fisher information matrix. 

\begin{corollary}
Let $\sigma(\underline{\theta})$, denote a family of density matrices parametrized by $\underline{\theta}$ as before, then   
\be \label{good-bad}
	\xi(\sigma) \leq \tilde{\xi}(\sigma).
\ee
Moreover, this immediately implies that
\be
	\|Cov(\underline{\hat{\theta}},\underline{\hat{\theta}})\|_{2 \raw 2} \geq \xi(\sigma)^{-1} \geq \tilde{\xi}(\sigma)^{-1}.
\ee
\end{corollary}

\proof{ The inequality (\ref{good-bad}) is a direct consequence of the min-max inequality,  \cite{Boyd}. Note, that after  appropriate reformulations we have 
\begin{eqnarray}
\xi(\sigma) &=& \max_{\{E^i\}} \min_{X}  \obra{X}\left(\sum_i \frac{\oproj{E^i}}{\tr{[E^i \sigma]}} \right) \oket{X} \nonumber \\
		&\leq& \min_{X} \max_{\{E^i\}}  \obra{X}\left(\sum_i \frac{\oproj{E^i}}{\tr{[E^i \sigma]}} \right) \oket{X} = \tilde{\xi}(\sigma),
\end{eqnarray}
for  traceless Hermitian and normalized $X$. The first inequality is the min-max inequality. In lemma \ref{chi2-dev} we have shown that the inner 
optimization over the POVM $\{E^i\}$ gives rise to the quantum Fisher information matrix with $\Omega_\sigma^B$.  \qed }

\vspace{0.5cm}
We think that the approach of minimizing the largest error determined in terms of the operator norm of the covariance matrix does provide a more suitable approach for determining  the optimal measurement in parameter estimation. Note, that the optimal measurement (\ref{optPOVM}) that maximizes the minimal eigenvalue of the classical Fisher information matrix is topographically complete. In turn, the measurement that is optimal for the quantum Fisher information (\ref{GoodMeasure}) is not. 
 
\section{Discussion}
We have studied the problem of hypothesis testing and goodness of fit testing of density matrices, and have focused on the $\chi^2$ test. This provides a clear, simple and flexible framework for testing whether a given density matrix is produced by a certain experimental setup, and allows to define confidence intervals that are independent of the particular system under consideration. We were also able to characterize divergence rates $\xi(\sigma)$ by doing an optimization over all possible POVM measurements maximizing the information, and proved that $2/3\leq \xi\leq 1$. This allowed to prove that, if we were sampling from a different density matrix $\rho$ instead of $\sigma$, that this would be detected in a number of measurements proportional to $d/(\xi(\sigma)\|\rho-\sigma\|^2)$ with $d$ the dimension of the Hilbert space. Furthermore, we showed that this measurement is both optimal from the point of view of Pitman and Bahadur efficiency.

We did not consider the question of entangled measurements on different copies. It would indeed be very interesting to understand, whether the approach take in \cite{Hayashi2} can be related to the scenario when different copies of the same state are considered. It is conceivable, that this approach could give rise to an optimal entangled measurement strategy when the contraint of i.i.d measurements is lifted; this will be discussed in future work.

Moreover, we have introduced a novel optimality criterium for quantum parameter estimation when multiple parameters need to be determined. We found that the largest possible error, as measured by the operator norm of the covariance matrix is lower bounded by the inverse of the divergence rate $\xi(\sigma)$, when all free parameters are estimated. The measurement that achieves the bound in the asymptotic limit is identical to the optimal POVM for quantum goodness of fit testing. 

We have only considered the case when all free parameters of the density matrix need to be estimated. In the general scenario, the dependence of the density matrix on the parameters may be more complicated. The investigation of this optimality criterium in the general scenario will be presented in a future publication. 

From a more philosophical point of view, the topic of hypothesis testing forces us to rethink what it means for a quantity to be physical and what not. For example, the expectation value of an observable is not observable, but can only be sampled. The resulting fluctuations are an entire part of doing an experiment, and if an experiment would report frequencies that are too close or too far from the expected ones, then such an experiment can be categorized as suspicious. The only thing that is physical are the frequencies by which certain measurement outcomes are obtained, and the only goal of quantum mechanics is the  prediction of those frequencies.

\section*{Acknowledgements}
We would like to thank Koenraad Audenaert for the help in formulating the proof of Lemma 5. This work was supported by the EU Strep project QUEVADIS, the ERC grant QUERG,  and the FWF SFB grants FoQuS and ViCoM. K.T. is grateful for the support from the Institute for Quantum Information and Matter, a NSF Physics Frontiers Center with support of the Gordon and Betty Moore Foundation (Grants No. PHY- 0803371 and PHY-1125565). An important part of this work was done at Stony Brook.

\end{document}